\documentclass{article}
\usepackage{spconf,amsmath,graphicx}
\usepackage{cite,multirow}
\usepackage{xcolor}

\usepackage{amssymb}
\usepackage{pifont}
\newcommand{\cmark}{\ding{51}}%
\newcommand{\xmark}{\ding{55}}%

\title{DATA AUGMENTATION WITH SIGNAL COMPANDING FOR DETECTION OF\\ LOGICAL ACCESS ATTACKS}

\name{Rohan Kumar Das$^{*}$, Jichen Yang$^{*}$ and Haizhou Li
\thanks{This research work is partially supported by Programmatic Grant No. A1687b0033 from the Singapore Government's Research, Innovation and Enterprise 2020 plan (Advanced Manufacturing and Engineering domain), Human-Robot Interaction Phase 1 (Grant No. 192 25 00054) by the National Research Foundation, Prime Minister's Office, Singapore under the National Robotics Programme. This work is also part of a collaboration with Kriston AI Lab, China in 2020.
*Corresponding Author}
}
\address{Department of Electrical and Computer Engineering, \\National University of Singapore, Singapore}
\begin{document}
%


\maketitle
\begin{abstract}

The recent advances in voice conversion (VC) and text-to-speech (TTS) make it possible to produce natural sounding speech that poses threat to automatic speaker verification (ASV) systems. To this end, research on spoofing countermeasures has gained attention to protect ASV systems from such attacks. While the advanced spoofing countermeasures are able to detect known nature of spoofing attacks, they are not that effective under unknown attacks. In this work, we propose a novel data augmentation technique using a-law and mu-law based signal companding. We believe that the proposed method has an edge over traditional data augmentation by adding small perturbation or quantization noise. The studies are conducted on ASVspoof 2019 logical access corpus using light convolutional neural network based system. We find that the proposed data augmentation technique based on signal companding outperforms the state-of-the-art spoofing countermeasures showing ability to handle unknown nature of attacks.  

\end{abstract}
\section{Introduction}
\label{sec:intro}

Automatic speaker verification (ASV) systems are used for a wide range of application services in the recent years~\cite{sv_debut,Das2016,SpeechMarker}. At the same time, spoofing attacks to these systems have become a concern as they are vulnerable to such attacks~\cite{Li2016_spoof_TD,bib:Attacker_overview2020}. In general, spoofing attacks are categorized into four major classes, which are impersonation, replay, voice conversion (VC) and text-to-speech synthesis (TTS) attacks~\cite{spoof_review}. The recent advances in VC and TTS technologies have produced not only high quality natural sounding speech~\cite{vcc2020summary}, but also show potential threat to ASV systems~\cite{Obama2018,vcc2020objective}.

The community driven ASVspoof\footnote{http://www.asvspoof.org/} challenge series promotes research on spoofing countermeasures using a benchmark corpus across research groups from last couple of editions. The third edition ASVspoof 2019 focuses on detection of logical access and physical access attacks in two separate tracks~\cite{ASVsppof2019_paper}. The logical access attacks are derived using the latest VC and TTS techniques, whereas the physical access attacks are created using replay samples in a simulated setup. In this work, we focus on the detection of logical access attacks as they show an imminent threat for unknown nature of attacks~\cite{ASVsppof2019_paper,vcc2020objective,Gen_CM_rkd}.

Literature shows that most of the novel explorations on spoofing countermeasures are based either on front-end handcrafted features or classifiers. Among these linear frequency cepstral coefficients (LFCC), subband spectral flux coefficients and spectral centroid frequency coefficients~\cite{ASVspoof2015_lfcc}, cochlear filter cepstral coefficient and instantaneous frequency (CFCCIF)~\cite{PatelINTERSPEECH2015} are few promising front-ends that proved effective in the first edition of ASVspoof 2015 to detect logical access attacks. Later, the constant-Q cepstral coefficients (CQCC)~\cite{CQCC_odyssey2016} derived from long-term constant-Q transform (CQT) emerged as a promising front-end that led to proposal of several handcrafted features along that direction~\cite{taslp_cmc,tifs_subband,elsevier_icqcc,rkd_is2019}. In the recent years, robust deep learning classifiers such as squeeze excitation residual networks~\cite{chengilaiITNTEERSPEECH2019,Alam_ASRU} and end-to-end systems with light convolutional neural networks (LCNN)~\cite{galinaITNTEERSPEECH2019,Yang2019} are found to be effective for detection of spoofing attacks.

Besides the front-end handcrafted features and robust classifiers, several studies are  focused on data augmentation for improving the performance against identifying unknown nature of attacks. The authors of~\cite{dataAug_Ming2017} carried out data augmentation by using parametric sound reverberator and phase shifter on the bonafide speech examples to simulate unseen conditions for replay speech. They extended their work for data augmentation by speed perturbation using bonafide and replay speech in the latest ASVspoof 2019 challenge to have effective detection of replay attacks~\cite{dataAug_Ming2019}. In~\cite{dataAug_apsipa19,dataAug_CSL}, vocal tract length perturbation is used apart from speed perturbation that boosted the performance for replay attack detection.

We find that although data augmentation has been used for handling replay or physical access attacks, it has not been much explored for dealing with logical access attacks. One of the reasons behind this may be due to the fact that replay attacks are affected by background acoustic conditions and therefore simulated conditions for data augmentation help to identify unknown nature of replay attacks. On the other hand, logical access attacks derived using VC and TTS may not be that useful to detect with traditional data augmentation as it may affect the artifacts discriminating bonafide and synthetic speech. Therefore, we believe methods not affecting acoustic properties may be useful for data augmentation for identifying unknown nature of logical access attacks on the evaluation set. In this work, we propose a novel data augmentation using signal companding techniques based on a-law and mu-law for detection of logical access attacks on ASVspoof 2019 corpus. We also compare our proposed approach with some of the existing data augmentation methods for comparison.

The remainder of the paper is organized as follows. Section~\ref{secii} discusses the proposed data augmentation based on signal companding. Section~\ref{seciii} describes the experiments conducted in the current study. The results and analysis are reported in Section~\ref{seciv}. Finally, Section~\ref{conc} concludes the paper.

\section{Signal Companding based Data Augmentation}
\label{secii}

In this work, we consider a novel way of performing data augmentation using signal companding techniques. Such methods compress and then expand the signals. The use of companding is popular for signals with a large dynamic range to be transmitted over facilities that have a smaller dynamic range capability. It is widely used in case of telephony speech and many other audio applications. We consider a-law and mu-law based signal companding methods that are two popular standard versions of G.711\footnote{https://www.itu.int/rec/T-REC-G.711} narrowband audio codec from ITU-T. Next, we discuss them in the following subsections. 

\subsection{a-law}

The a-law based companding technique is used in European 8-bit PCM digital communications as per ITU-T standards. It  reduces the dynamic range of the signal, thereby increasing the coding efficiency and resulting in a signal-to-distortion ratio that is superior to that obtained by linear encoding for a given number of bits. For a given signal $x$, the a-law encoding is performed as follows
\begin{equation}
F_a(x)={\text {sgn}}(x)
\begin{cases}
\frac{A|x|}{1+\ln(A)}, |x|<\frac{1}{A}\\
\frac{1+\ln(A|x|)}{1+\ln(A)}, \frac{1}{A}\leq|x| \leq 1
\end{cases}
\end{equation}
where the compression parameter $A=$86.5 on European standards and ${\text {sgn}}(x)$ is the sign function. The a-law expansion is then performed as follows
\begin{equation}
{F_a}^{-1}(y)={\text {sgn}}(y)
\begin{cases}
\frac{|y|(1+\ln(A))}{A}, |y|<\frac{1}{1+\ln(A)}\\
\frac{\exp(|y|(1+\ln(A))-1)}{A}, \frac{1}{1+\ln(A)}\leq|y| \leq 1
\end{cases}
\end{equation}
\subsection{mu-law}
The mu-law is another kind of standard companding technique, which is used in North America and Japan as per ITU-T standards. It provides a slightly larger dynamic range than a-law based approach. For a given signal $x$, the mu-law encoding is performed as follows
\begin{equation}
    F_\mu(x)={\text {sgn}}(x)\frac{\ln(1+\mu|x|)}{\ln(1+\mu)}, -1\leq x \leq 1
\end{equation}
where $\mu$ is the compression parameter, which equals to 255 in North American and Japanese standards. The mu-law expansion is then performed as follows
\begin{equation}
    {F_\mu}^{-1}(y)={\text {sgn}}(y)(1/\mu)((1+\mu)^{|y|}-1), -1\leq y \leq 1
\end{equation}

We use the above discussed a-law and mu-law based companding techniques to increase the number of training examples for data augmentation to build robust spoofing countermeasure model for detection of unknown nature of logical access attacks. As this method does not require any additional database for data augmentation, it has an edge over some of the traditional data augmentation methods, where external datasets with noise or room reverberation are used. 

\section{Experiments}
\label{seciii}

In this section, we discuss the experiments conducted for the current work. The details of the corpus and experimental setup are mentioned in the following subsections.

\subsection{Corpus}


The ASVspoof 2019 logical access corpus\footnote{https://datashare.is.ed.ac.uk/handle/10283/3336} is used for the studies in this work~\cite{ASVsppof2019_paper}. The database has three subsets that are train, development and evaluation set. The bonafide examples of the corpus are taken from VCTK\footnote{http://dx.doi.org/10.7488/ds/1994} corpus. There are 46 male and 61 female speakers totalling 107 speakers in the corpus. The three subsets of the corpus do not have any speaker overlap. In addition, the spoofed examples of evaluation set are derived using different TTS and VC methods from those used in the train and development set. The evaluation protocol of ASVspoof 2019 considers tandem detection cost function (t-DCF) and equal error rate (EER) as performance metrics for reporting the results~\cite{Kinnunen2018_TDCF}. It is noted that the ASV-centric t-DCF measure is obtained by combining our spoofing countermeasure scores with the ASV scores given along with ASVspoof 2019 corpus. A summary of the ASVspoof 2019 logical access corpus is presented in Table~\ref{tab:la2019}.

 \begin{table}[!t]
 \begin{center}
 \caption{ASVspoof 2019 logical access corpus summary.}
 \label{tab:la2019}
 \vspace{-2mm}
\resizebox{8.7cm}{!}{
 \begin{tabular}{|c|c|c|c|c|}
 \hline {\bf{Subset}}       &{\bf{\#Male}} &{\bf{\#Female}}  &{\bf{\#Bonafide}} &{\bf{\#Spoofed}} \\
 \hline
 \hline Train     & 8   & 12     & 2,580    & 22,800 \\
 \hline Development  & 4   & 6     & 2,548    & 22,296 \\
 \hline Evaluation   & 21  & 27     & 7,355    & 63,882 \\
 \hline
 \end{tabular}
 }
 \end{center}
 \vspace{-8mm}
 \end{table}

\subsection{Experimental setup}

In this study, we use long-term CQT based log power spectrum (LPS) as input to the LCNN system similar to that in~\cite{SJTU_ASVspoof2019}. The static dimension of LPS is 84, where the number of octaves is 7 and the number of frequency bins in every octaves is 12. In order to extract the LPS of fixed dimension, we set the length as 550 frames either by padding or cropping that makes input feature of 84$\times$550 for each example. The architecture of the LCNN system implemented using PyTorch toolkit follows our previous work~\cite{LCNN_IS2020}. It is noted that optimal number of layers and nodes are obtained on the development set.

Both a-law and mu-law based signal companding methods are used for augmentation of training data to train new models for detection of logical access attacks.
In other words, we increase the amount of training data by three times with examples derived using a-law and mu-law signal companding. We also perform traditional data augmentation by small amount of noise addition for comparison. The NoiseX-92 database~\cite{NoiseX92} is used for comparative noise data augmentation studies. We use four noise categories that are cafe, street, volvo and white noise with 20 dB SNR for data augmentation. 

\section{Results and Analysis}
\label{seciv}

We now analyze and discuss the experimental results. For brevity, we refer the proposed data augmentation with signal companding as DASC in short.


\subsection{Effect of signal companding data augmentation}

We are first interested in a comparison between DASC and a baseline without data augmentation. Second, we would like to compare our results with the two baselines of ASVspoof 2019 implemented using CQCC and LFCC based front-end with Gaussian mixture models (GMM).

Table~\ref{tab:la2019dev} reports the comparison for with and without DASC, as well as ASVspoof 2019 baselines. It is noted that the attacks in the evaluation set of ASVspoof 2019 logical access corpus are derived using a wide range of unseen VC and TTS methods compared to that used in the development set, which results in a performance difference from the development set. In addition, the robustness of a spoofing countermeasure depends on its effectiveness for detection of the unknown nature of attacks on the evaluation set.

We find from Table~\ref{tab:la2019dev} that our CQT-LCNN baseline system without any data augmentation performs much better than the ASVspoof 2019 challenge baselines projecting it as a strong state-of-the-art system. Further, when we apply DASC, we obtain 1.16\% absolute improvement in EER on the evaluation set. This validates the DASC idea for detection of unknown logical access attacks. 

We now compare the performance for models with and without DASC when signal companding is applied on the test set. Table~\ref{tab:la2019a-mulaw} shows the results for this comparison, which reveals that the result of our baseline without DASC degrades slightly in EER when a-law and mu-law based signal companding technique applied on the test set. However, the performance of proposed DASC does not have much difference from its original performance under this scenario. This further strengthens the effectiveness of the proposed DASC.

\begin{table}[!t]
\begin{center}
\caption{Performance of CQT-LCNN system with and without DASC, along with  the challenge baselines in ASVspoof 2019 logical access database.}
\vspace{-2mm}
\label{tab:la2019dev}
\resizebox{8.7cm}{!}{
\begin{tabular}{|c|c|c|c|c|}
\hline
{\bf{Proposed }} & \multicolumn{2}{|c|}{\bf{Development Set}}& \multicolumn{2}{|c|}{\bf{Evaluation Set}}\\
\cline{2-5}
 {\bf{ DASC}}    &{\bf{t-DCF}}           &{\bf{EER (\%)}}      &{\bf{t-DCF}}           &{\bf{EER (\%)}}   \\ \hline
\hline \xmark        &  0.023         &0.77     &     0.129      &4.29 \\
{\cmark}     &{0.028 }           &{ 0.86}     &{\bf 0.094}           &{\bf 3.13}  \\
\hline
\hline
\multicolumn{5}{|c|}{\bf{Baselines of ASVspoof 2019 Challenge~\cite{ASVsppof2019_paper}}}\\
\hline \hline
CQCC-GMM & 0.0123& 0.43  &0.2366 & 9.57 \\
LFCC-GMM &0.0663 & 2.71  & 0.2116& 8.09\\
\hline
\end{tabular}
}
\end{center}
\vspace{-8mm}
\end{table}

\begin{table}[!t]
\begin{center}
\caption{Performance of CQT-LCNN system with and without DASC considering a-law and mu-law companding applied on ASVspoof 2019 logical access evaluation set.}
\label{tab:la2019a-mulaw}
\vspace{-2mm}
\resizebox{8.7cm}{!}{
\begin{tabular}{|c|c|c|c|c|}
\hline
{\bf{Test Data}} & \multicolumn{2}{|c|}{\bf{Model without DASC}}& \multicolumn{2}{|c|}{\bf{ Model with DASC}}\\
  \cline{2-5}
{\bf{Companding}} & {\bf{t-DCF}} & {\bf{EER (\%)}} & {\bf{t-DCF}} & {\bf{EER (\%)}} \\ \hline
\hline
a-law & 0.130 &  4.89 & {\bf 0.097} & {\bf 3.12}\\ 
mu-law & 0.125 & 4.56 & {\bf 0.095} & {\bf 3.08}\\
 \hline
\end{tabular}
}
\end{center}
\vspace{-8mm}
\end{table}

\subsection{Comparison with traditional data augmentation}

We would like to further compare our proposed DASC method to some traditional way of performing data augmentation. In this regard, we consider four noises to augment the training data for creating new models for anti-spoofing. The results are reported in Table~\ref{tab:la2019noise}. It is observed that the traditional way of performing data augmentation does not help to improve the performance in case of detection of logical access attacks. The DASC system performs much better than all the considered noise cases.

\begin{table}[!t]
\begin{center}
\caption{Performance comparison of proposed DASC with traditional noise (20 dB) based data augmentation on ASVspoof 2019 logical access database. We test all models on standard ASVspoof 2019 logical access evaluation set.}
\label{tab:la2019noise}
\vspace{-2mm}
\resizebox{8.7cm}{!}{
\begin{tabular}{|c|c|c|c|c|}
\hline
{\bf{Model with Data }} & \multicolumn{2}{|c|}{\bf{Development Set}}& \multicolumn{2}{|c|}{\bf{Evaluation Set}}\\
\cline{2-5}
 {\bf{Augmentation}}    &{\bf{t-DCF}}           &{\bf{EER (\%)}}      &{\bf{t-DCF}}           &{\bf{EER (\%)}}   \\ \hline\hline
{ DASC }     &{\bf 0.028}           &{\bf 0.86}     &{\bf 0.094}           &{\bf 3.13}  \\
\hline\hline
Cafe Noise  &  0.106  & 3.79  & 0.245  & 8.22 \\
Street Noise & 0.130  & 4.67  & 0.247  & 9.20 \\
Volvo Noise & 0.090   & 3.23  & 0.186  & 7.06 \\
White Noise & 0.125   & 4.12 & 0.282  & 10.63 \\
 \hline
\end{tabular}
}
\end{center}
\vspace{-8mm}
\end{table}

We further extend the studies to perform testing under noisy condition. The same four categories of noise are added to the evaluation set data. We evaluate the system without data augmentation, data augmentation with noise, and DASC and report in Table~\ref{tab:la2019noiseTest}. We note that the spoofing countermeasure models with noise based data augmentation considers the respective model matched to that of the test noise case. It is observed from Table~\ref{tab:la2019noiseTest} that under noisy testing scenario, the noise based data augmentation model performs better than the baseline model in majority cases without any data augmentation. This indicates the robustness of the noise based data augmentation model towards noisy testing. 

Further, the proposed signal companding based data augmentation model achieves improved performance over the traditional noise based data augmentation for noisy scenario expect for white noise case. This shows that our proposed system even performs effectively against unknown noisy data showing its robustness. However, the case of white noise did not show this trend. It may be because white noise has a flat power spectra, which may lead to a better detection under matched case when white noise based data augmentation is performed making the scenario more predictable.

\begin{table}[!t]
\begin{center}
\caption{Performance comparison of proposed DASC with traditional noise (20 dB) based data augmentation on ASVspoof 2019 logical access database. We test all models on noise-added ASVspoof 2019 logical access evaluation set.}
\label{tab:la2019noiseTest}
\vspace{-2mm}
\resizebox{8.7cm}{!}{
\begin{tabular}{|c|c|c|c|c|c|c|}
\hline
{\bf{Noisy}} & \multicolumn{6}{|c|}{\bf{Model}}\\ \cline{2-7}
{\bf{Test}} & \multicolumn{2}{|c|}{\bf{Without Data}}& \multicolumn{2}{|c|}{\bf{With Noise based }}& \multicolumn{2}{|c|}{\bf{With Proposed}}\\
 {\bf{ Case}} & \multicolumn{2}{|c|}{\bf{Augmentation}}& \multicolumn{2}{|c|}{\bf{Data Augmentation}}& \multicolumn{2}{|c|}{\bf{DASC}}\\
 \cline{2-7}
{\bf{(20 dB)}}    &{\bf{t-DCF}}           &{\bf{EER (\%)}}      &{\bf{t-DCF}}           &{\bf{EER (\%)}} &{\bf{t-DCF}}           &{\bf{EER (\%)}}    \\ \hline
\hline
Cafe &  0.261 &	9.26  & 0.229 &	8.93  &{\bf{ 0.157}}  &  {\bf{5.62}}\\
Street & 0.406 & 14.89   & 0.282 &	11.12 & {\bf{0.279}}  & 	{\bf{9.92}}  \\
Volvo &  0.157  &	5.78  &0.181  &	6.81  &{\bf{0.143}}   &	{\bf{5.62}}  \\
White & 0.457	&16.75 &0.283	 &11.42 &  0.450 &	14.33  \\
 \hline
\end{tabular}
}
\end{center}
\vspace{-8mm}
\end{table}

\subsection{Comparison with other known single systems}

In this subsection, we would like to compare the proposed DASC system with various single systems available on evaluation set of ASVspoof 2019 logical access corpus. We consider some of the top performing systems of ASVspoof 2019 challenge as well as recent works published post challenge. These single systems use different front-end  features and classifiers. 

We consider novel front-ends single frequency cepstral coefficients (SFCC), zero time windowing cepstral coefficients (ZTWCC) and instantaneous frequency cepstral coefficients (IFCC) based systems reported in~\cite{knrkrajuITNTEERSPEECH2019}. Similarly, several deep learning systems like LCNN, residual network (ResNet) and deep neural network (DNN) are also considered that use different inputs such as constant-Q statistics-plus principal information coefficients (CQSPIC), mel frequency cepstral coefficient (MFCC), LFCC, CQCC, feature genuinization (FG), LPS of discrete Fourier transform (DFT) and fast Fourier transform (FFT)~\cite{rkd_ASRU2019,moustafaITNTEERSPEECH2019,galinaITNTEERSPEECH2019}.

Table~\ref{la2019comsys} shows the performance comparison of our proposed system with signal companding to other known single system results on ASVspoof 2019 evaluation set. We find that the proposed system outperforms all other single system results in terms of both the performance metrics t-DCF and EER. This projects our proposed system as a robust anti-spoofing system to handle unknown nature of logical access attacks derived using TTS and VC.

\begin{table}[!t]
 \begin{center}
 \caption{Performance comparison of the proposed spoofing countermeasure using DASC with some known single systems on ASVspoof 2019 logical access evaluation set.}
 \vspace{2mm}
 \label{la2019comsys}
 \begin{tabular}{|c|c|c|}
 \hline {\bf{System }}                                            &{\bf{t-DCF}}                     &{\bf{EER (\%)}}    \\  \hline
 \hline 
  \hline SFFCC-GMM~\cite{knrkrajuITNTEERSPEECH2019}                                 &0.323                     &13.97         \\
  \hline ZTWCC-GMM~\cite{knrkrajuITNTEERSPEECH2019}                                 &0.141                     &6.13         \\
 \hline IFCC-GMM~\cite{knrkrajuITNTEERSPEECH2019}                                  &0.357                     &15.59         \\
 \hline LFCC-DNN~\cite{rkd_ASRU2019}                                         &0.234                     &9.65         \\ 
 \hline CQCC-DNN~\cite{rkd_ASRU2019}                                         &0.308                     &12.79        \\
 \hline MFCC-ResNet~\cite{moustafaITNTEERSPEECH2019}                               &0.204                     &9.33         \\
  \hline CQCC-ResNet~\cite{moustafaITNTEERSPEECH2019}                               &0.217                     &7.69         \\
 \hline LPS-DFT-ResNet~\cite{moustafaITNTEERSPEECH2019}                               &0.274                     &9.68         \\
 \hline CQSPIC-DNN~\cite{rkd_ASRU2019}                                           &0.183                     &7.81         \\
 \hline CQSPIC-GMM~\cite{rkd_ASRU2019}                                           &0.164                     &7.74         \\
 \hline LFCC-LCNN~\cite{galinaITNTEERSPEECH2019}                                   &0.100                     &5.06         \\
 \hline LPS-FFT-LCNN~\cite{galinaITNTEERSPEECH2019}                                    &0.103                     &4.53         \\ 
 \hline { {FG-LCNN}~\cite{LCNN_IS2020}}                                             &{ {0.102}}   &{ {4.07}}         \\\hline
 \hline {\bf {Proposed DASC}}                                             &{\bf {0.094}}   &{\bf {3.13}}         \\
 \hline
 \end{tabular}
 \end{center}
 \vspace{-6mm}
 \end{table}

\section{Conclusion}
\label{conc}

This work proposes a novel data augmentation technique using a-law and mu-law based signal companding for detection of logical access attacks. The studies conducted on ASVspoof 2019 logical access corpus reveal that the proposed data augmentation is able to detect the unknown nature of attacks on the evaluation set more effectively than that without data augmentation. In addition, the comparison to traditional noise based data augmentation method shows that the proposed method is more effective. The proposed system with signal companding based data augmentation also outperforms existing state-of-the-art single spoofing countermeasure systems on ASVspoof 2019 logical access corpus. The future work will focus on extending the proposed data augmentation technique for other speech and audio processing applications. 

\newpage
\footnotesize
\bibliographystyle{IEEEbib}
\bibliography{MyReferences_new}

\end{document}